\documentclass[useAMS]{mn2e}
\usepackage{graphicx}
\newcommand\fu{4U~0115+634}
\title[Detection of a variable QPO at $\sim$41 mHz in 4U~0115+634]{Detection of a variable QPO at $\sim$41 mHz in the Be/X-ray transient pulsar 4U~0115+634}

\author[Dugair et al.]{Moti R. Dugair$^{1}$\thanks{dugair\_moti@rediffmail.com}, Gaurava K. Jaisawal$^{2}$, Sachindra Naik$^{2}$ and S. N. A. Jaaffrey$^{1}$\\
$^1$Astronomy and Astrophysics Laboratory, Department of Physics, Mohanlal Sukhadia University, Udaipur - 313001, Rajasthan, India\\
$^2$Astronomy and Astrophysics Division, Physical Research Laboratory, Navrangapura, Ahmedabad - 380009, Gujarat, India\\}

\begin{document}

\date{\bf Accepted for publication in MNRAS}

\maketitle

\begin{abstract}

We report the detection of  quasi-periodic oscillation (QPO) at $\sim$41 mHz in the transient 
high-mass Be/X-ray binary pulsar 4U~0115+634 using data from the $Rossi~X-ray~Timing~Explorer 
(RXTE)$ observatory. The observations used in the present work were carried out during X-ray 
outbursts in 1999 March-April, 2004 September-October and 2008 March-April. This frequency of
the newly detected QPO was found to vary in 27-46 mHz range. This $\sim$41 mHz QPO was 
detected in four of the 36 pointed $RXTE$ Proportional Counter Array (PCA) observations
during 1999 outburst where as during 2004 and 2008 outbursts, it was detected in four and 
three times out of 33 and 26 observations, respectively. Though QPOs at $\sim$2 mHz, and 
$\sim$62 mHz were reported earlier, the $\sim$41 mHz QPO and its first harmonic were 
detected for the first time in this pulsar. There are three $RXTE$/PCA observations where 
multiple QPOs were detected in the power density spectrum of 4U~0115+634. Simultaneous 
presence of multiple QPOs is rarely seen in accretion powered X-ray pulsars. Spectral 
analysis of all the pointed $RXTE$/PCA observations revealed that the 3-30 keV energy 
spectrum was well described by Negative and Positive Power-law with EXponential cutoff 
(NPEX) continuum model along with interstellar absorption and cyclotron absorption 
components. During the three X-ray outbursts, however, no systematic variation in any 
of the spectral parameters other than the earlier reported anti-correlation between 
cyclotron absorption energy and luminosity was seen. Presence of any systematic variation 
of QPO frequency and rms of QPO with source flux were also investigated yielding negative 
results.

\end{abstract}


\section{Introduction}

Be/X-ray binaries (BeXBs) represent the largest sub-class of high mass X-ray binaries (HMXBs). 
BeXBs are known to consist of a neutron star and a non-supergiant OB star (luminosity class of 
III, IV or V) showing Balmer lines in the emission spectrum (Reig 2011). The optical companion 
in BeXB systems (Be star) is still on the main sequence. The neutron star in these systems is generally in moderate eccentric (e$\ge$0.3) and wide orbit with orbital period in range of 
16-400 days. Mass transfer from the optical companion to the neutron star takes place through 
the equatorial circumstellar disk around the Be star. Though the neutron star spends most of 
the time away from the Be circumstellar disk, during the periastron passage, abrupt mass 
accretion takes place resulting in periodic X-ray outbursts, known as Type~I outbursts. The
peak luminosity during these periodic outbursts is moderate ($L_X \leq 10^{35-37}$ erg 
s$^{-1}$ ; Stella, White \& Rosner 1986). The duration of these outbursts is in the range 
of a few days to a few tens of days. The BeXBs occasionally show giant outbursts, known as 
Type~II outbursts. These outbursts are caused by the enhanced episodic outflow of the Be 
star. The peak luminosity during Type~II outbursts is estimated 
to be as high as 10$^{38}$ erg s$^{-1}$ or more (Stella, White \& Rosner 1986; Negueruela et 
al. 1998). These outbursts are irregular and not linked with the binary orbit of the system. 
The neutron stars in most of the BeXB systems are found to be accretion powered pulsars. The 
pulse period of these pulsars is in the range of a few seconds to several hundreds of seconds. 
The X-ray spectrum of these pulsars are found to be hard. For a brief review on the temporal 
and spectral properties of transient BeXB pulsars, refer to Paul \& Naik (2011).

\begin{figure*}
\centering
\includegraphics[height=6.5in, width=2.3in, angle=-90]{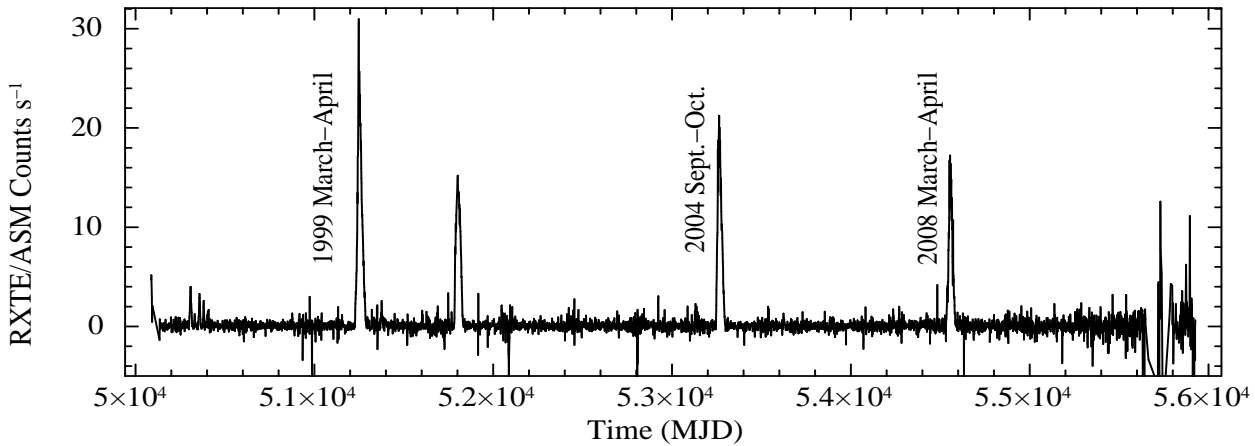}
\caption{$RXTE$/ASM one-day averaged light curve of the transient pulsar 4U~0115+634 in 
1.5--12 keV energy range from 1996 January 5 (MJD 50087) to 2011 December 31 (MJD 55926).
During entire observing period of $RXTE$, only four major outbursts were detected in the 
ASM light curve. $RXTE$/PCA observations during 1999, 2004 and 2008 outbursts were analyzed 
to investigate the QPOs features in the pulsar.}
\label{fg1}
\end{figure*}

The Be/X-ray transient pulsars 4U~0115+634 was discovered during the $UHURU$ satellite survey
(Giacconi et al. 1972; Forman et al. 1978). Using the survey data, the pulsation of the transient
pulsar was first estimated to be 3.6 s (Cominsky et al. 1978). A few years after the discovery, a 
new outburst was detected with $SAS~3$ during 1977 December to 1978 January. Using the precise 
position of the pulsar obtained from the $SAS~3$, $Ariel~V$ and $HEAO-1$ observations (Cominsky 
et al. 1978; Johnston et al. 1978), the optical counterpart was identified to be a strongly 
reddened Be star with a visual magnitude $V \sim 15.5$ and named as V635~Cas (Johns et al. 1978;
Hutchings \& Crampton 1981). Timing observations of the pulsar with $SAS~3$ were used to estimate 
orbital parameters of the binary system such as orbital period P$_{orb}$ = 24.31$\pm$0.02 days,
eccentricity of the orbit $e$ = 0.34$\pm$0.01, projected semi-major axis $a_{x}sin~i$ = 140$\pm$0.16
lt-sec, $\omega$ = 47$^\circ$.66$\pm$0.$^{\circ}$17, mass function $f(M) \approx$ 5 M$_\odot$ and
time of periastron passage $\tau$ = 2443540.95$\pm$0.01 days (Rappaport et al.1978). The distance 
to the pulsar 4U~0115+63 was estimated as 7-8 kpc (Negueruela \& Okazaki 2001). A detailed study
of optical and X-ray emissions from the binary system showed that the Type~II outbursts in this
pulsar were detected at an interval of about every three years (Whitlock et al. 1989). This was
explained in terms of instabilities in the circumstellar disk due to radiative warping (Negueruela 
et al. 2001). The continuum spectrum of the pulsar was well described by a power-law model with 
high energy cut-off (Rose et al. 1979; White et al. 1983). A cyclotron resonance scattering feature 
(CRSF) at $\sim$23 keV was first detected in the $HEAO-1$ A4 spectra of pulsar during an outburst 
(Wheaton et al. 1979) which was later suggested to be the first harmonic resonance, with the 
fundamental resonance at $\sim$11 keV (White et al. 1983). This was later confirmed with $Ginga$ 
observation of the pulsar (Nagase et al. 1991). Second, third and fourth harmonics of the CRSF 
are later detected in the pulsar spectrum from $RXTE$ and $BeppoSAX$ observations (Heindl et al. 
1999; Santangelo et al. 1999). A close monitoring of the 1999 March-April outburst of the pulsar 
with $RXTE$ confirmed the luminosity dependence of the CRSF in \fu~ which was interpreted as a 
result of the decrease in height of the accretion column, in response to a decrease in the mass
accretion rate (Nakajima et al. 2006). Similar anti-correlation between the CRSF energy and
X-ray luminosity has been seen in other binary X-ray pulsars such as A0535+262, Her~X-1 etc. 
(Terada et al. (2006); Mihara et al. (2007); Enoto et al. (2008) and references therein). 
$HEAO~1$ observation of the pulsar \fu, during an outburst reported possible presence of QPOs 
at 62 mHz (Soong \& Swank 1989). Apart from the 62 mHz QPO, another low frequency QPO at 2 mHz 
were detected in $RXTE$ observations of the pulsar during 1999 March-April outburst (Heindl et 
al. 1999). These QPOs were explained as due to the obscuration of the neutron star by hot 
matter in the accretion disk. The blobs of hot matter in the inner accretion disk are 
understood to be because of the in-homogeneities caused by the interactions of the neutron 
star magnetosphere and the accretion disk.

In the present work, we have investigated the timing and spectral properties of the transient 
X-ray pulsar 4U~0115+634 using observations made with the $RXTE$/PCA and report the detection of
variable QPO features detected during three X-ray outbursts in 1999 March, 2004 September \& 
2008 April.

\begin{figure*}
\vskip 5.8cm
\includegraphics{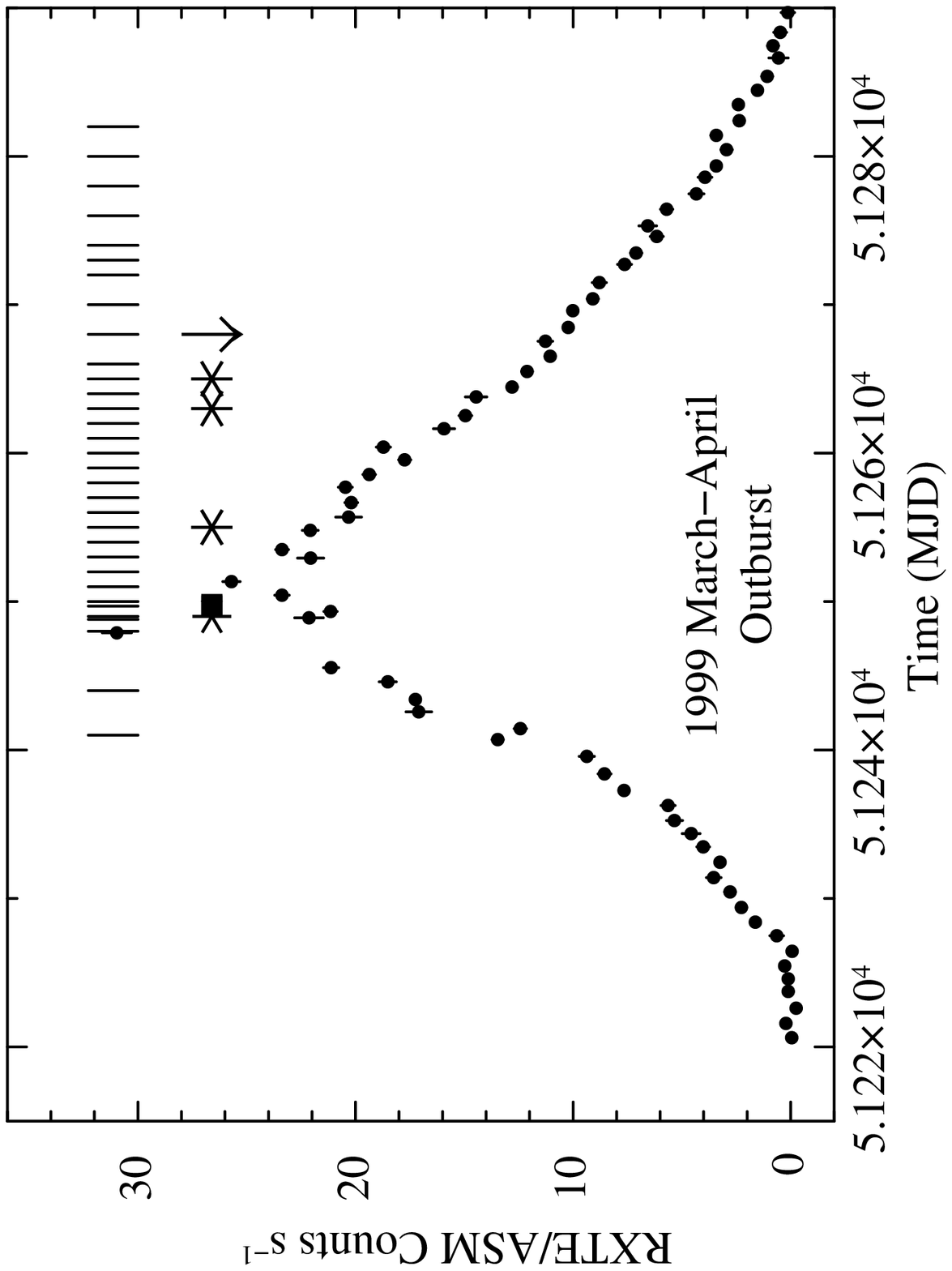}
\includegraphics{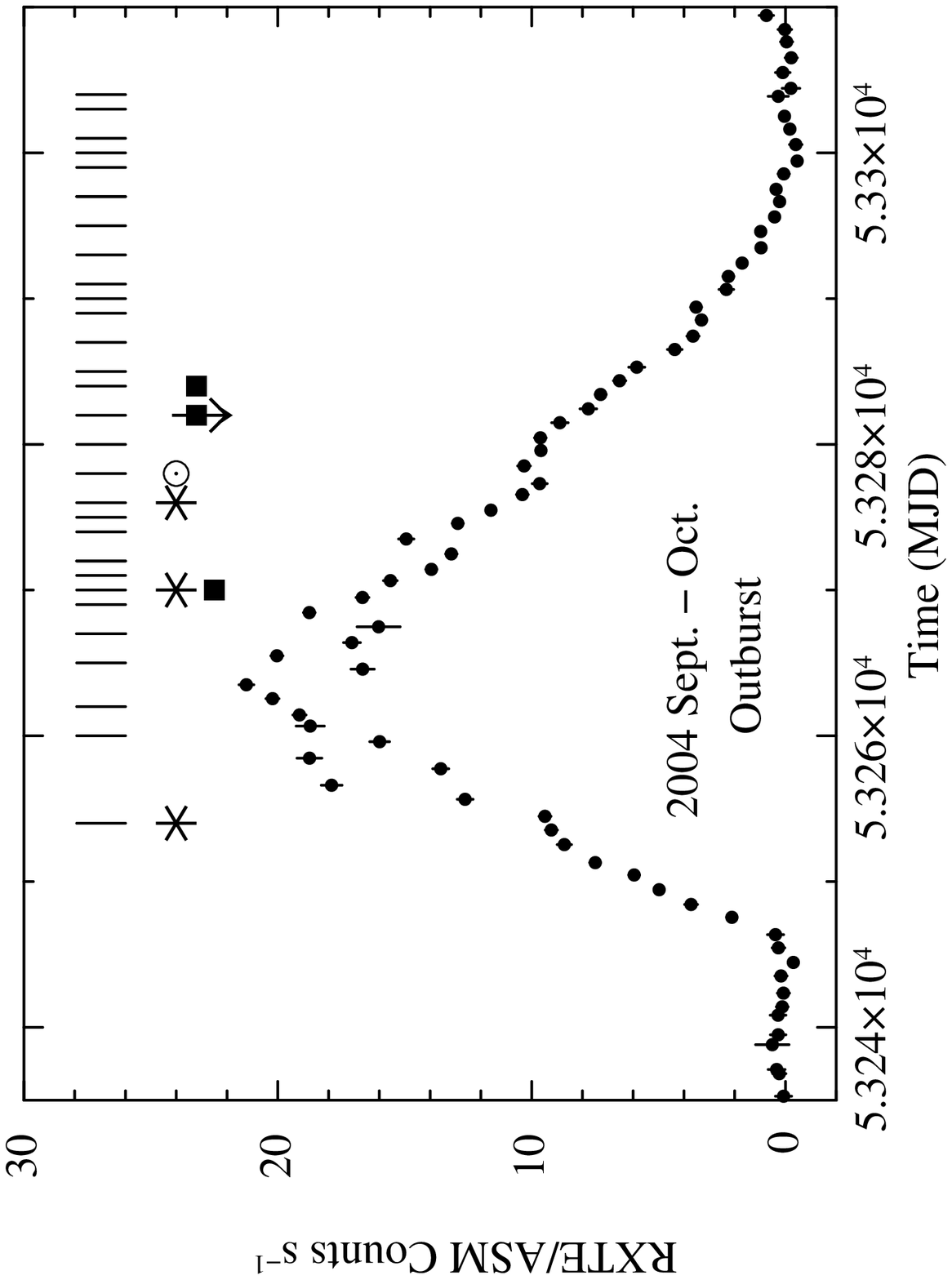}
\includegraphics{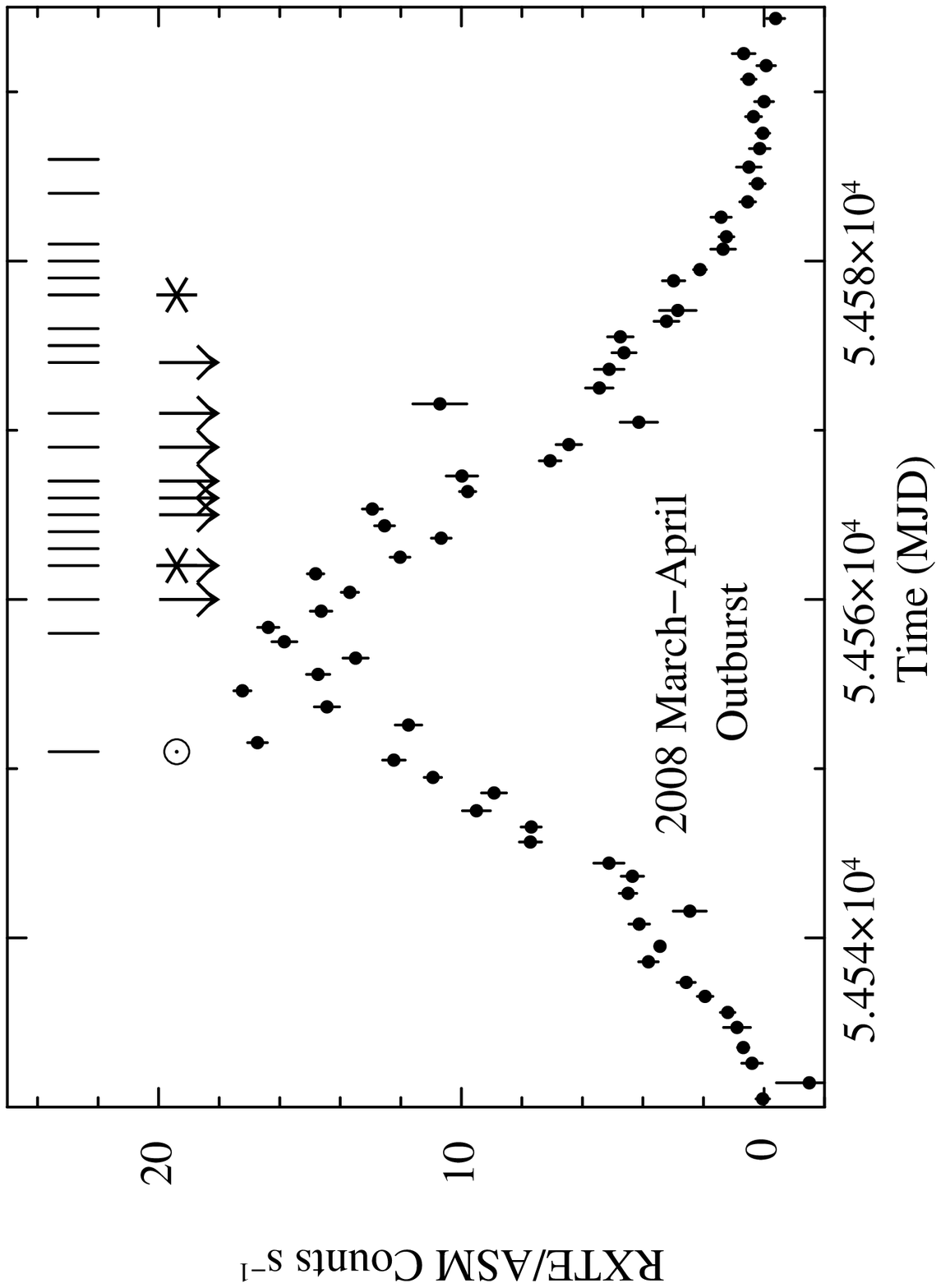}
\caption{One-day averaged $RXTE$/ASM light curves of the transient Be/X-ray binary
pulsar 4U~0115+634 during 1999 March-April (left panel), 2004 September-October (middle
panel) and 2008 March-April (right panel) outbursts. The vertical lines at the top of
each panels indicate the $RXTE$/PCA pointed observations of the pulsar. The solid squares,
vertical arrows and asterisks in each panels indicate the observations in which 
QPOs at $\sim$2 mHz, $\sim$62 mHz and the newly detected QPO at $\sim$41 mHz were 
detected, respectively. The ``$\sun$'' symbols in second and third panels show the
presence of newly detected $\sim$41 mHz QPO and its first harmonic in the power
density spectra of the pulsar. There are a few observations during which presence 
of multiple QPOs in the PDS was also seen and marked in the figure.}
\label{fg2}
\end{figure*}

\section{Observation and Data Analysis}

$RXTE$ was launched on 1995 December 31 with the main objective of timing studies of celestial 
X-ray sources. It made great contributions towards the understanding of high energy astrophysics 
by means of its unrivaled timing resolution. We have used the $RXTE$ observations of transient
BeXB pulsar 4U~0115+634 during three of its Type~II outbursts. $RXTE$, which is
now decommissioned, had three sets of major instruments. The all sky monitor (ASM) was sensitive 
in 1.5-12 keV energy range (Levine et. al. 1996). The PCA, which was consisting of five Xenon 
filled proportional counter detectors, was sensitive in 2-60 keV energy range. The effective area, 
energy resolution and time resolution of PCA were $\sim$6500 cm$^2$ at 6 keV,  $\leq$18\% at 
6 keV and 1 $\mu$s, respectively. A detailed description of the PCA instrument can be found in 
Jahoda et al. (1996). The third instrument, High Energy Timing Experiment (HEXTE) was operating 
in 15-250 keV energy range (Rothschild et al. 1998). 

\subsection{Timing Analysis}

The $RXTE$/ASM light curve of 4U~0115+634 from 1996 January to 2011 December is shown in
Figure~\ref{fg1}. During above duration i.e. entire life time of the $RXTE$ observatory,
only four major outbursts were detected in the pulsar. Out of the four outbursts, during
2000 outburst, the pulsar was monitored only three times and hence the data were not used in our
present analysis. However, the pulsar was extensively monitored during the other three outbursts. 
For our timing analysis, we used data from all PCA observations during the 1999 March-April, 
2004 September-October and 2008 March-April outbursts (as marked in Figure~\ref{fg1}).  
In this work, we analyzed a total of 95 publicly available $RXTE$/PCA observations (36 during 
1999 March-April outburst, 33 during 2004 September-October outburst and 26 during 2008 
March-April outburst). A brief log of the observations is given in Table~\ref{log}. In this 
work, we used PCA data from these observations to study the evolution of QPO. Data reduction 
was carried out by using the software package FTOOLS where as data analysis was done by 
using the HEAsoft package (ver 6.11). 

\begin{table}
\centering
\caption{List of RXTE/PCA observations during 1999, 2004 and 2008 outburst.}
\begin{tabular}{llcc}
\hline
Year of &Observation  &No. of  &Total\\
Outburst     &Series   &IDs  &Exposure (ks) \\
\hline
1999  &P40051   &15   &74.6 \\
      &P40070   &5    &117.2 \\
      &P40411   &16   &23.4 \\
\\      
2004  &P90014   &16   &40.8 \\
      &P90089   &17   &65\\
\\
2008  &P93032   &26   &228.9\\      
\hline
\end{tabular}
\label{log}
\end{table}

Using the standard~1 mode PCA data, we extracted light curves in 2-60 keV energy range with 
time resolution of 0.125 s from all the $RXTE$ pointed observations during the 1999, 2004 and 
2008 outbursts. While extracting the source light curves, we found that different number of 
PCUs were on during different dates of observations. Corresponding background light curves
were created by using the modeled background event data (as provided by the instrument team).
Background subtracted average source count rates at peak of the 1999, 2004 and 2008
outbursts were found to be $\sim$1238 counts s$^{-1}$, $\sim$1170 counts s$^{-1}$ and 
$\sim$1115 counts s$^{-1}$ per PCU, respectively. Data from all available PCUs were used
in our analysis. Power-density spectra (PDS) were generated from each of the light curves 
obtained from the $RXTE$ observations of the pulsar during 1999, 2004 and 2008 outbursts 
by using the FTOOLs package {\it powspec}. To detect the presence/absence of QPO, PDS were 
created for small segments of duration 512 s and averaged over the total number of available 
segments in the individual light curves. The PDS were normalized to subtract the white 
noise level. These normalized PDS were used to estimate the squared rms fractional 
variability by integrating over a certain frequency range. All the PDS were then examined 
for the presence/absence of QPOs in a wide frequency range ($\sim$1 mHz to $\sim$1 Hz). 

\begin{figure}
\centering
\includegraphics[height=2.8in, width=2.4in, angle=-90]{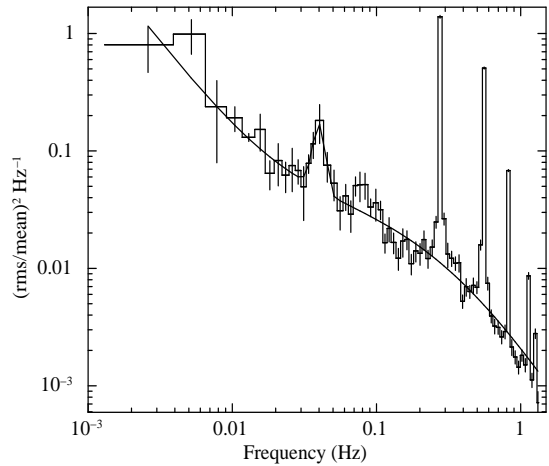}
\caption{The power density spectrum of 4U~0115+63 obtained from the $RXTE$/PCA observation 
on 1999 March 25. The presence of newly detected $\sim$41 mHz QPO is clearly seen in the PDS. 
The 3.6 s pulsations and its harmonics are seen at higher frequencies. The solid line in the 
figure represents the fitted model comprising of a power-law continuum, a Lorentzian function 
and a Gaussian function.} 
\label{1999_qpo}
\end{figure}

\begin{figure}
\centering
\includegraphics[height=2.8in, width=2.4in, angle=-90]{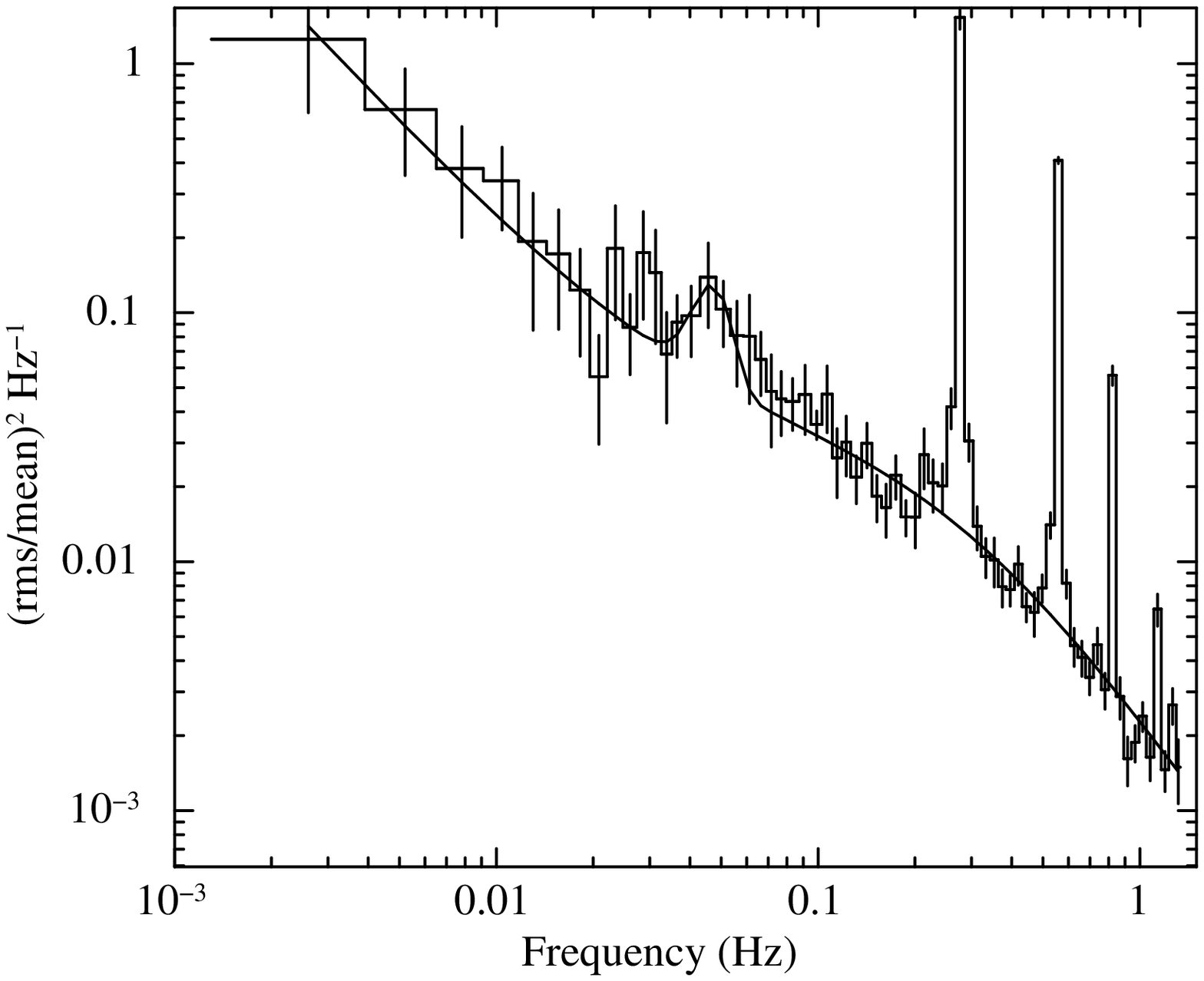}
\caption{The power density spectrum of 4U~0115+63 obtained from the $RXTE$/PCA observation on 2004
September 6. The presence of newly detected $\sim$41 mHz QPO is clearly seen in the PDS. The 3.6 s 
pulsations and its harmonics are seen at higher frequencies. The solid line in the figure
represents the fitted model comprising of a power-law continuum, a Lorentzian function and 
a Gaussian function.} 
\label{2004_qpo}
\end{figure}

\begin{figure}
\centering
\includegraphics[height=2.8in, width=2.4in, angle=-90]{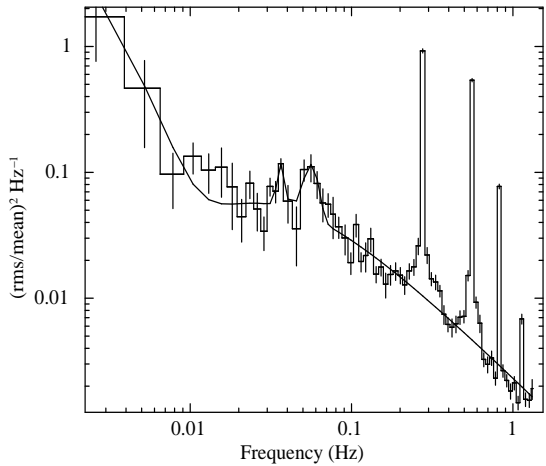}
\caption{The power density spectra of 4U~0115+63 obtained from the $RXTE$/PCA observation on 2008
April 6. The presence of newly detected $\sim$41 mHz QPO is seen in the PDS along with the previously detected $\sim$62 mHz QPO. The 3.6 s pulsations are seen at higher frequencies. The solid line in the figure represents the fitted model comprising of a power-law continuum, a Lorentzian function and a Gaussian function.} 
\label{2008_qpo2}
\end{figure}

\begin{figure}
\centering
\includegraphics[height=2.8in, width=2.4in, angle=-90]{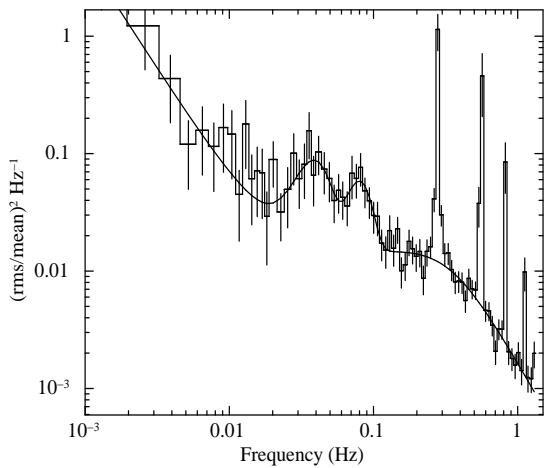}
\caption{The power density spectra of 4U~0115+63 obtained from the $RXTE$/PCA observation on 2008
March 26. The presence of newly detected $\sim$41 mHz QPO is clearly seen in the PDS along with 
its first harmonic. The 3.6 s pulsations are seen at higher frequencies. The solid line in 
the figure represents the fitted model comprising of a power-law continuum, a Lorentzian function
and a Gaussian function.} 
\label{2008_qpoH}
\end{figure}

We found that the 3.6 s regular pulsations of the pulsar and its harmonics were present 
in the PDS obtained from all the $RXTE$/PCA observations. Apart from these pulsations 
and corresponding harmonics, the PDS from 72 $RXTE$/PCA observations during the 
three outbursts of the pulsar were featureless. However, in some cases, very prominent 
QPO features at $\sim$41 mHz were detected though the strength was variable during other 
observations. To estimate the frequency of the QPO, we fitted the PDS (for 1999 March 25
observation) with two continuum components such as a power law and a Lorentzian. The peaks
corresponding to the spin frequency of the pulsar and its harmonics were ignored while 
fitting the PDS continuum. The fit clearly showed the presence of a Gaussian feature at 
$\sim$41 mHz (shown in Figures~\ref{1999_qpo} \& \ref{2004_qpo}). This $\sim$41 mHz QPO 
in the transient pulsar 4U~0115+634 is detected for the first time here though QPOs at 
$\sim$2 mHz (Heindl et al. 1999) and the possibility of the presence of QPO at $\sim$62 
mHz (Soong \& Swank 1989) were reported earlier. We obtained the ratio of the amplitude 
and uncertainty of the Gaussian component and found the QPO feature is more than 3$\sigma$
detection level. It may be noted that the exposure of individual $RXTE$/PCA observations 
were not very large (poor signal-to-noise ratio) which enhanced the error in the Gaussian 
component and degraded the significance level of the QPO feature. However, the feature can be 
clearly seen in the PDS as shown in one of the observations (Figure~\ref{1999_qpo}). 
While careful examining this $\sim$41 mHz QPO during the three outbursts, it is found that 
the QPO frequency is variable in the range of 27-46 mHz and the significance level of the 
QPO was always in the range of 3$\sigma$-8$\sigma$. Apart from this newly detected $\sim$41 
mHz QPO, the other two QPOs were also detected in the PDS of some of the $RXTE$/PCA pointed observations of the pulsar. While carefully investigating all the $RXTE$ observations 
of \fu~ during 1999, 2004 and 2008 outbursts, the presence of multiple QPOs were detected 
in the PDS of a few epochs. Out of 95 pointed $RXTE$ observations, multiple QPOs were seen
only in three occasions viz. 2004 September 22 (QPOs at $\sim$2 mHz and $\sim$41 mHz), 2004
October 4 (QPOs at $\sim$2 mHz and $\sim$62 mHz) and 2008 April 6 (QPOs at $\sim$41 mHz and 
$\sim$62 mHz). Apart from the presence of multiple QPOs at three epochs, we also detected
$\sim$41 mHz QPO and its first harmonic on two occasions such as 2004 September 30 and
2008 March 26. Figure~\ref{2008_qpo2} shows presence of $\sim$41 mHz and $\sim$62 mHz QPOs
on 2008 April 6 whereas Figure~\ref{2008_qpoH} shows the presence of $\sim$41 mHz QPO and
its first harmonic in the PDS of 2008 March 26 observation. These findings i.e. presence
of multiple QPOs at a particular epoch and the detection of QPOs and its harmonics are
very rarely seen in accretion powered X-ray pulsars.

We calculated the quality factor $Q$ (QPO central frequency $\nu$/FWHM (full width at half
maximum)) of the all three QPOs detected during 1999, 2004 and 2008 outbursts and found to 
be in the range of 3-12 apart from three cases where it is in the range of 15-20. The estimated
value of $Q$ is comparable to that in other high mass X-ray binaries (4-10). Conventionally, 
a local maximum in the PDS is interpreted as a QPO when the quality factor $Q$ is more than 2 
(van der Klis 2000). The high value of Q ($\geq$2) for the peak at $\sim$41 mHz in the PDS of 
\fu~  also confirms the feature to be a QPO. The rms values of the detected QPOs in the 
$RXTE$/PCA observations were estimated and found to be in the range of 4\% to 8\% except the 
two observation of 1999 outburst where the rms value of 2 mHz QPO is quite high $\sim$17\%.
Comparisons between the QPO parameters with the flux of the pulsar are given in next section. 
One-day averaged 1.5-12.0 keV $RXTE$/ASM light curves of the pulsar during the 1999, 2004 and 
2008 outbursts are shown in Figure~\ref{fg2}. In each panel, the day of $RXTE$ observations 
used in present work are marked with vertical lines at the top. Solid squares, asterisks and 
arrow marks in each panel represent the observations in which QPOs at $\sim$2 mHz, $\sim$41 mHz 
and $\sim$62 mHz are detected, respectively. In second and third panels, the $\sun$ marks
represent the presence of newly detected $\sim$41 mHz QPO and its first harmonic in the power
density spectrum of the pulsar. Out of a total of 95 $RXTE$ observations, we found QPOs in 
23 observations.

\subsection{Spectral Analysis}

To investigate the changes in spectral parameters during the 1999, 2004 and 2008 outbursts
and then compare these changes with the QPO parameters, we carried out spectral analysis of
the available $RXTE$/PCA observations of the pulsar during above outbursts. As majority
of the observations are not long enough, data from HEXTE instruments were not used in our
analysis. Out of 33, 33 and 26 pointed $RXTE$ observations of the pulsar during 1999, 2004 
and 2008 outbursts, spectral analysis were carried out for 29, 27 and 24 observations,
respectively. Standard-2 mode PCA data from all the $RXTE$/PCA observations were used 
to create source spectrum. Standard procedures were applied for the data selection, background 
estimation and response matrix generation (Naik \& Paul 2003). Data from available PCUs were 
added together and response matrices were generated accordingly. We restricted our analysis 
to the 3-30 keV energy range. Though the $RXTE$ observations of the pulsar during 1999 
outburst were analyzed to investigate the luminosity dependence of Cyclotron Resonance 
Scattering Features (CRSF; Nakajima et al. 2006), we used these observations in the context 
of understanding the change in spectral parameters and also QPO parameters with source flux. 
We used same spectral continuum model as used by Nakajima et al. (2006) such as the Negative 
and Positive power-law with EXponential cutoff (NPEX) model along with the CRSF while fitting 
the pulsar spectra during all three outbursts. The NPEX model is known to be the approximation 
of the unsaturated thermally comptonized plasma (Makishima et al. 1999). As described by 
Nakajima et al. (2006), a CRSF at $\sim$11 keV and its harmonic are seen in the 3-30 keV 
energy spectra of the pulsar. The CRSF energy was found to vary inversely with the pulsar 
luminosity which was interpreted as a result of decrease in the accretion column height 
due to the decrease in mass accretion rate (Nakajima et al. 2006). 

\begin{figure}
\centering
\includegraphics[height=3.in, width=2.4in, angle=-90]{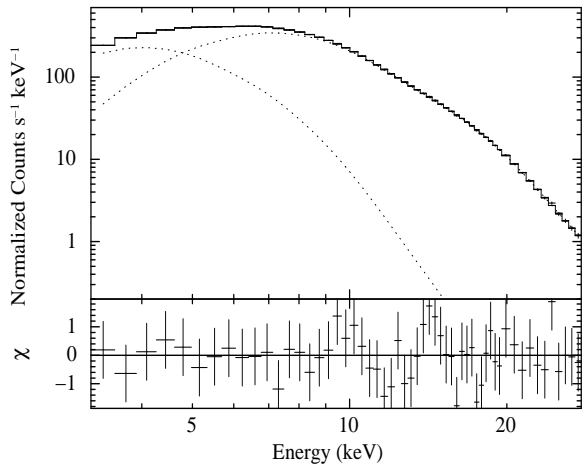}
\caption{The energy spectrum of 4U~0115+63 obtained from the $RXTE$/PCA observation on 
1999 March 14, along with the best-fitting model comprising of NPEX continuum model and
CRSF. The bottom panel shows the contributions of the residuals to the $\chi^2$ for each 
energy bin for the best-fitting NPEX continuum model.}
\label{spec}
\end{figure}

\begin{figure*}
\centering
\includegraphics[height=6.5in, width=2.6in, angle=-90]{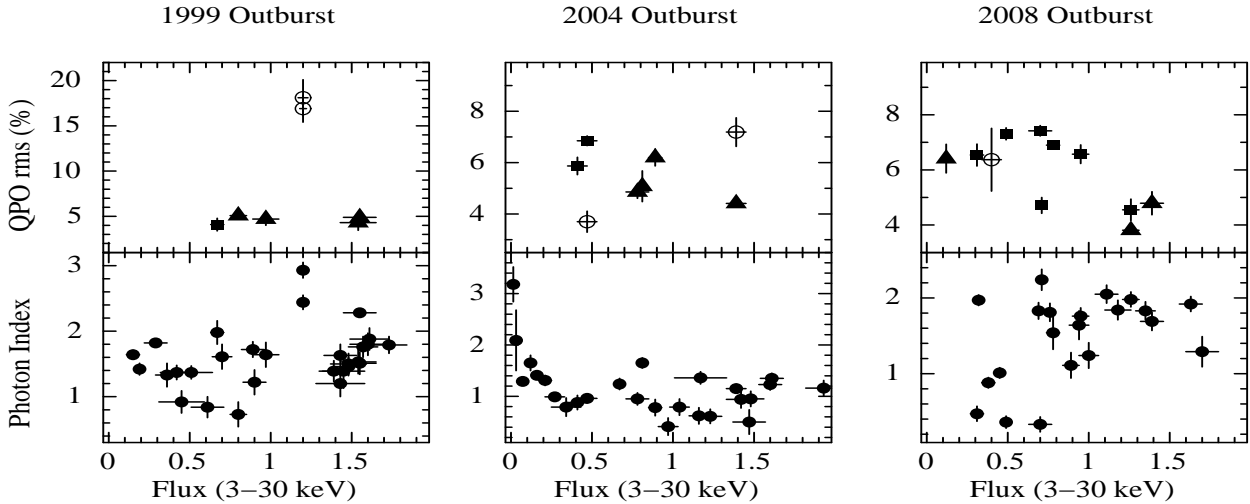}
\caption{Change in the QPO rms (\%) and power-law photon index with source flux 
(in $10^{-8}$ erg cm$^{-2}$ s$^{-1}$ units) in 3-30 keV energy range. The open circles,
solid triangles and solid squares in top three panels represent the rms values for QPOs at
$\sim$2 mHz, $\sim$41 mHz and $\sim$62 mHz, respectively. The left, middle 
and right panels are for 1999, 2004 and 2008 outbursts, respectively.}
\label{fl_qpo}
\end{figure*}

As a detailed study on the luminosity dependence of CRSF is already done, we attempted to
investigate the change in other parameters with the source flux during above three outbursts. 
A representative energy spectrum of the pulsar obtained from the $RXTE$ observation on 1999 
March 14 is shown in Figure~\ref{spec} along with the best-fitting model and the residuals. 
We estimated source flux in 3-30 keV energy range along with the power-law photon index, 
high energy cutoff and equivalent hydrogen column density ($N_H$) for all the 80 observations
during three outbursts. Considering the energy range of spectral fitting (i.e. 3-30 keV), 
the value of estimated hydrogen column density may not be accurate enough to draw any 
reliable conclusion. Random variation was seen in the values of power-law photon index 
with 3-30 keV source flux during 1999 and 2008 outbursts. Similar variation in the values
of photon index with the pulsar luminosity was reported earlier during the 1999 outburst
earlier (see Table~4 of Nakajima et al. 2006). However, during 2004 outburst, the photon 
index was found to be high at low source flux level. This should be noted here that during
2004 outburst, the pulsar was monitored with $RXTE$/PCA even after the outburst (quiescent
phase). The high value of power law photon index during quiescent phase suggests that the 
pulsar spectrum was relatively hard compared to that during the outburst phase. In 
Figure~\ref{fl_qpo}, we plotted the values of power law photon index and the rms (\%) of 
QPOs with the estimated source flux in 3-30 keV energy range during all three outbursts. 
It was found that the rms (\%) of the QPO was variable in 3-8\% range during all three
outbursts except $\sim$17\% rms value of 2 mHz strong QPOs of 1999 outburst. However, there 
was no systematic variation in the values of QPO rms during all three outbursts.

\section{Discussion}

The frequency of the QPOs detected in accretion powered X-ray pulsars generally falls 
in the range of $\sim$1 mHz to $\sim$40 Hz (Psaltis 2006). Among these binary pulsars, 
the QPO features are detected more in transient sources compared to the persistent ones.
In most of the transient Be/X-ray binary pulsars, the detected QPOs are found to be 
transient in nature. For example - in case of KS~1947+300, the QPO feature was not seen 
in the data during 2000 and 2002 outbursts whereas it appeared at the end of the 2001 
outburst (James et al. 2010). In 4U~0115+634, the QPO features were not detected in all
of the $RXTE$ observations during 1999, 2004 and 2008 outbursts (present work). Transient 
HMXB pulsars in which QPOs have been detected are EXO~2030+375 (Angelini et al. 1989), 
4U~0115+63 (Soong \& Swank 1989), V0332+53 (Takeshima et al. 1994), A0535+262 (Finger 
et al. 1996), SAX~J2103.5+4545 (Inam et al. 2004), XTE~J1858+034 (Mukherjee et al. 2006),
XTE~J0111.2-7317 (Kaur et al. 2007), MAXI~J1409-619 (Kaur et al. 2010),  KS~1947+300 
(James et al. 2010), 4U~1901+03 (James et al. 2011), 1A~1118-61 (Devasia et al. 2011a),  
and GX~301-4 (Devasia et al. 2011b).  However, there are only very few cases of 
accretion-powered X-ray pulsar such as SMC~X-1 (Wojdowski et al. 1998), Her~X-1, 
4U~1626-67, LMC~X-4 (see Table~1 of Shirakawa \& Lai 2002) and 4U~0115+63 (present work) 
where multiple QPOs are seen. Another unique thing we detected in this work is the 
presence of multiple QPOs simultaneously in the same observation at three occasions. 

QPOs in accretion powered X-ray binary pulsars are thought to be due to the motion of
the inhomogeneously distributed matter in the inner part of the accretion disk. The
detection of QPOs, therefore, provides useful information on the radius of the inner
accretion disk and the interaction between accretion disk and the neutron star. 
Several models have been proposed to explain the QPO features in the PDS of accretion 
powered X-ray pulsars. According to the magnetospheric beat frequency model, the observed 
QPO frequency ($\nu_{QPO}$) represent the beat between the coherent spin frequency of 
the pulsar ($\nu_s$) and the Keplerian frequency ($\nu_K$) at the inner disk radius
i.e. $\nu_{QPO}$ = $\nu_{K}$(r$_{in}$) - $\nu_{s}$, at the magnetospheric boundary of 
the pulsar (Alpar \& Shaham, 1985; Lamb et.al. 1985). According to this model, interactions
between the neutron star magnetosphere and its accretion disk result in fluctuations in the
plasma density at the inner edge of the disk which rotates with the local Keplerian 
frequency $\nu_K$(r$_{in})$. As the magnetic field lines of the neutron star rotate with 
its spin frequency $\nu_s$, the fluctuation in plasma reappears with a frequency that is 
the beat frequency as given above. In Keplerian-frequency model, on the other hand, the 
QPOs arise because of the modulation of the X-rays by inhomogeniously distributed matter 
in the inner accretion disk at the Keplerian frequency (van der Klis et al. 1987). 
According to this model, the observed QPO frequency is same as the frequency of the 
Keplerian motion at the inner accretion disk (i.e. $\nu_{QPO}$ = $\nu_{K}$(r$_{in}$)). 
When the spin frequency of the pulsar is higher than the Keplerian frequency at the 
inner edge of the accretion disk, mass accretion on to the neutron star is stopped 
at the magnetospheric boundary by centrifugal inhibition of accretion (Stella, White 
\& Rosner 1986) resulting in the onset of propeller effect. Keplerian-frequency model, 
therefore, can be applicable only when the QPO frequency is above the neutron star spin 
frequency, as seen in transient BeXB pulsars such as EXO~2030+375 (Angelini, Stella \& 
Parmar 1989), A0535+262 (Finger et al. 1996), XTE~J0111.2−7317 (Kaur et al. 2007) etc. 
However, in case of 4U~0115+634, frequency of all three QPOs (earlier reported QPOs at 
$\sim$2 mHz and $\sim$62 mHz and newly detected QPO at $\sim$41 mHz) are found to be 
less than the spin frequency of the pulsar. Therefore, the Keplerian-frequency model 
is not suitable to explain the presence of QPO in the transient pulsar 4U~0115+634.

Third model proposed to explain the low-frequency QPO features in strongly
magnetized ($\sim$10$^{12}$ Gauss) accretion powered X-ray pulsars is the 
magnetic disk precession model (Shirakawa \& Lai 2002). According to this 
model, the presence of QPOs in binary pulsars is due to the magnetically 
driven disk warping/precession near the inner edge of the disk at the 
magnetospheric boundary. The magnetic torques due to interactions of the 
stellar field and the induced electric currents in the disk is responsible 
for warping and precession of the disk. This model was used to explain the 
mHz QPO detected in the accretion-powered pulsar 4U~1626-67 along with 
several others (Shirakawa \& Lai 2002). This model was also attempted 
to explain the QPOs in the transient pulsar 4U~0115+634 in terms of the 
presence of magnetically driven disk warping/precession around the neutron 
star.

\section{Conclusion}

We reported the detection of a $\sim$41 mHz QPO for the first time in the transient pulsar 
4U~0115+634. These newly detected QPOs were present more frequently in decline phase of 
outbursts as compare to the rising phase (Figure~\ref{fg2}). These $\sim$41 mHz QPOs were 
found to be variable in 27-46 mHz frequency range. Apart from the newly detected $\sim$41 
mHz QPO, QPOs at $\sim$2 mHz and $\sim$62 mHz were also detected at several other occasions. 
However, there was no correlation between the QPO parameters and source flux during above 
three outbursts in the pulsar. In this work, we have detected the presence of multiple 
QPOs in the PDS of the pulsar only on three occasions viz. on 2004 September 22 (QPOs at 
$\sim$2 mHz and $\sim$41 mHz), 2004 October 04 (QPOs at $\sim$2 mHz and $\sim$62 mHz) and
2008 April 06 (QPOs at $\sim$41 mHz and $\sim$62 mHz) whereas $\sim$41 mHz QPO and its 
first harmonic were detected on 2004 September 30 and 2008 March 26. The presence of multiple 
QPOs and QPOs with its harmonics in a single observation are very rarely seen in accretion 
powered X-ray pulsars.

\section*{Acknowledgments}

The authors would like to thank the anonymous referee for his/her useful comments and 
suggestions that improved the contents of the paper. MD would like to acknowledge the 
hospitality provided by the Physical Research Laboratory, Ahmedabad, India during his 
visit to carry out the present work. The work of MD was supported by University Grant
Comission, New­ Delhi, India under the financial support scheme RGN Fellowship (No.F.14­ 
2(SC)/2008(SA­III)). This research has made use of data obtained through the High Energy
Astrophysics Science Archive Research Center Online Service,, provided by the NASA/Goddard 
Space Flight Center.

\end{document}